\documentclass[aps,prb,twocolumn,showpacs,amsmath,amssymb]{revtex4}
\usepackage{graphicx}
\usepackage{bm}

\def\br{ \bm{r} }
\def\bk{ \bm{k} }

\begin{document}

\title{Tunneling into $d$-wave superconductors: Effects of interface spin-orbit coupling}

\author{S. Wu and K. V. Samokhin}

\affiliation{Department of Physics, Brock University, St. Catharines, Ontario L2S 3A1, Canada}

\date{\today}

\begin{abstract}
Tunneling conductance of a clean normal metal/$d$-wave superconductor junction is studied by using the extended
Blonder-Tinkham-Klapwijk formalism. We show that the conductance is significantly affected by the interface spin-orbit coupling of the Rashba type,
which is inevitably present due to the asymmetry of the junction.
\end{abstract}


\maketitle


High-temperature cuprate superconductors have remained at the forefront of
experimental and theoretical research for more than twenty years. Probing the order parameter structure
has been the subject of a particularly large effort. One of the most compelling pieces
of evidence for the $d$-wave symmetry of the order parameter comes from tunneling spectroscopy experiments.
The conductance spectrum of a junction between a normal metal and a high-$T_c$
superconductor exhibits strong dependence on the crystallographic orientation of the interface, see Refs.
\onlinecite{ABS-review-1} and \onlinecite{ABS-review-2} for a review. Its most prominent feature is the zero-bias conductance peaks (ZBCP) that can be attributed to the quasiparticle states with zero energy bound to the surface.\cite{Hu94,TanKash95prl,KTKK96}
Such states, called the Andreev bound states, exist if the quasiparticles experience a sign change of the order parameter upon reflection
from the interface, as was originally pointed out by Hu.\cite{Hu94}
The Andreev bound states and the associated low-energy features in the tunneling conductance have also been studied for other unconventional superconductors.\cite{ABS-review-1,YTK97,HonSig98,LS07,Inio07}

Most of the theoretical studies of the tunneling conductance in high-$T_c$ superconductors have used the
Blonder-Tinkham-Klapwijk (BTK) formalism,\cite{BTK82} extended to the $d$-wave case.
It has been known, however, that going beyond the BTK model produces important qualitative effects.
For instance, time-reversal symmetry can be spontaneously broken near the interface due to the formation of a subdominant order parameter,
leading to the splitting of the ZBCP even in zero external magnetic field,\cite{FRS97,TanKashi01}
while in the presence of the interface roughness, the ZBCP exist for all interface orientations.\cite{FRS97}
Even within the BTK framework, the tunneling conductance in the $d$-wave case turns out to be sensitive to the details of the interface barrier,
see, e.g., Ref. \onlinecite{ZhuWang97}, where the effects of ferromagnetic and Kondo-like scattering in the barrier were considered.

In this Letter we study the
effects of the spin-orbit coupling (SOC) localized near an interface between a normal metal and a $d$-wave superconductor.
Due to the fact that two sides of the junction have different crystal and electronic structure, the interface potential barrier is asymmetric, resulting in
the SOC of the type originally proposed by Rashba in Ref. \onlinecite{Rashba60} for semiconductor heterostructures.
We neglect disorder as well as the interface roughness and calculate the zero-temperatute tunneling conductance for different crystalline orientations
by generalizing the BTK formalism to include the Rashba interface SOC. Similar model was recently applied in Ref. \onlinecite{WuSam10} to a normal metal/$p$-wave
superconductor junction. Throughout the paper we use the units in which $\hbar=1$.

We consider the tunneling junction shown in Fig.~\ref{fig: model}. The interface is located at $x=0$ and is
characterized microscopically by a potential barrier which we describe by the following model:
\begin{equation}
\label{interface-model}
    U(x)=[U_{\textrm{0}}+U_{1}\bm{n}\cdot(\hat{\bm{\sigma}}\times\hat{\bk})]\delta(x).
\end{equation}
Here $\bm{n}\equiv\hat{\bm{x}}$ is the unit vector along the interface normal, $U_0$ and $U_1$ are the strengths of the spin-independent and the Rashba SOC contributions, respectively, $\hat{\bm{\sigma}}$ are the Pauli matrices, and $\hat{\bk}=-i\bm{\nabla}$.
The band dispersions are assumed to be parabolic, with the same effective masses $m$ and the Fermi energies $k_F^2/2m$ on both sides.

The quasiparticle wave function has four components, corresponding to the electron-hole and spin degrees of freedom, and can be found from the Bogoliubov-de Gennes (BdG) equations.\cite{DeGennes}
Assuming two-dimensional geometry, the SOC is diagonal in spin, and the BdG equations can be decoupled into two independent pairs of two-component
equations as follows:
\begin{equation}
\label{BdG-eqs-2}
    {\cal H}_\sigma\Psi(\br)=E\Psi(\br),
\end{equation}
where $\sigma=\pm$ for different spin orientations,
\begin{equation}
\label{H-BdG-sigma}
    {\cal H}_\sigma=\left(\begin{array}{cc}
    \displaystyle \hat\xi+U_{\sigma}(x) & \sigma\Delta(\hat{\bk},\br)\\
    \sigma\Delta^{\dag}(\hat{\bk},\br) & \displaystyle -\hat\xi-U_{\sigma}(x)
    \end{array}\right),
\end{equation}
$\hat\xi=\hat{\bk}^2/2m-\epsilon_F$, and $U_{\sigma}(x)=(U_{0}-\sigma U_{1}\hat{k}_{y})\delta (x)$.
The gap function is given by $\Delta(\hat{\bk},\br)=\Delta(\hat{\bk})\theta(x)$, where
$\theta(x)$ is the step function. The spin index in the off-diagonal elements of ${\cal H}_\sigma$ amounts to
an unimportant phase factor and can be dropped when calculating the tunneling conductance. We note that the values of the gap can be different for the transmitted
electron-like and hole-like quasiparticles, due to the anisotropy of the $d$-wave order parameter. We have
\begin{equation}
\label{Delta-pm}
    \Delta_{\pm}\equiv\Delta(\bk_{\pm})=\Delta_{0}\cos(2\theta\mp2\alpha),
\end{equation}
where $\alpha$ is the angle between the crystalline orientation and
$x$-axis, $\bk_\pm=k_F(\pm\cos\theta,\sin\theta)$, and $\Delta_{+}$ and $\Delta_{-}$ are the effective pair
potentials of electron and hole components, respectively. In the spirit of the BTK approach, self-conistency of the order parameter is neglected.

\begin{figure}
\includegraphics[width=10.5cm]{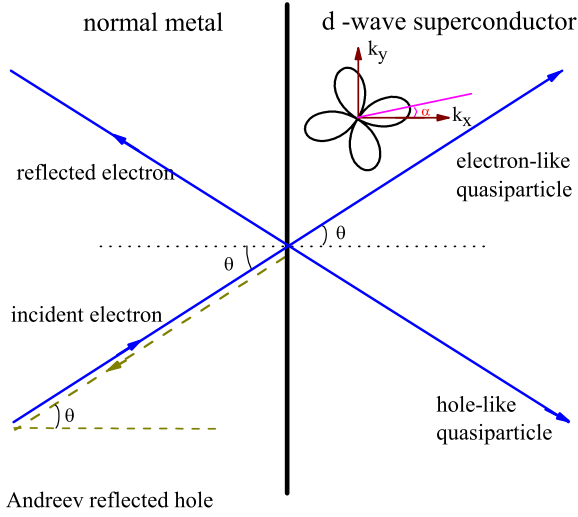}
\caption{(Color online) Schematic illustration of the quasiparticle
reflection and transmission processes at the interface. Also
shown is the $d$-wave order parameter profile and the angle $\alpha$
of the crystalline orientation with respect to $x$-axis.} \label{fig:
model}
\end{figure}

We assume that electrons are injected from the normal metal with
the excitation energy $E\geq 0$ and spin $\sigma$, at an angle
$\theta$ from the interface normal. The incident electrons are either normally reflected as electrons or
Andreev reflected as holes.\cite{And64} The momentum parallel to the
interface is conserved in the tunneling process and the solution of Eq. (\ref{BdG-eqs-2}) has the form
$\Psi(\br)=e^{ik_yy}\Psi(x)$, where
\begin{eqnarray}
\label{Psi_N}
    \Psi_N(x)&=&\left(\begin{array}{c}
    1\\ 0
    \end{array}\right)e^{ik_{F}\cos\theta\, x}
    +a_{\sigma}
    \left(\begin{array}{c}
    0\\ 1
    \end{array}\right)e^{ik_{F}\cos\theta\, x}
    \nonumber\\
    &&+b_{\sigma}
    \left(\begin{array}{c}
    1\\ 0
    \end{array}\right)e^{-ik_{F}\cos\theta\, x}
    \end{eqnarray}
on the normal side, and
\begin{eqnarray}
\label{Psi_S}
    \Psi_{S}(x)&=&c_{\sigma}
    \left(\begin{array}{c}
    u_{+}e^{i\phi_{+}}\\ v_{+}
    \end{array}\right)e^{ik_{F}\cos\theta\,x}\nonumber\\
    &&+d_{\sigma} \left(\begin{array}{c}
    v_{-} e^{i\phi_{-}}\\ u_{-}
    \end{array}\right)e^{-ik_{F}\cos\theta\,x}
\end{eqnarray}
on the superconducting side. Here $a_{\sigma}$ and $b_{\sigma}$ are the amplitudes of the Andreev and normal
reflection, respectively, and $c_{\sigma}$ and $d_{\sigma}$ are the transmission amplitudes. The quasiparticle amplitudes in the
superconducting region are given by
\begin{eqnarray}
    u_{\pm}=\frac{1}{\sqrt{2}}\sqrt{1+\frac{\Omega_{\pm}}{E}},
    \qquad v_{\pm}=\frac{1}{\sqrt{2}}\sqrt{1-\frac{\Omega_{\pm}}{E}},
\end{eqnarray}
where $\Omega_{\pm}=\sqrt{E^{2}-|\Delta_{\pm}|^{2}}$, with the phase factors $e^{i\phi_{\pm}}=\Delta_{\pm}/|\Delta_{\pm}|$. Note that, according to Eq. (\ref{Delta-pm}), $\phi_\pm=0$ or $\pi$.

All the reflection and transmission amplitudes in Eqs. (\ref{Psi_N}) and (\ref{Psi_S}) can be found from the
boundary conditions that follow from Eq. (\ref{interface-model}):
\begin{eqnarray*}
    \left.\begin{array}{l}
    \Psi_S(0^{+})=\Psi_N(0^{-}),\\ \\
        \Psi'_S(0^{+})-\Psi'_N(0^{-})=2m(U_0-\sigma U_{1}k_{F}\sin\theta)\Psi_N(0^{-}).\end{array}\right.
\end{eqnarray*}
In particular, for the reflection amplitudes we obtain:
\begin{equation}
    \left.\begin{array}{l}
    \displaystyle a_{\sigma}(E,\theta)=\frac{4
    \Gamma_{+}e^{-i\phi_{+}}}{(2+Z_{\sigma}^{2})\omega_{-}+2\omega_{+}},\\ \\
    \displaystyle
    b_{\sigma}(E,\theta)=\frac{(-2iZ_{\sigma}-Z_{\sigma}^{2})
    \omega_{-}}{(2+Z_{\sigma}^{2})\omega_{-}+2\omega_{+}},
    \end{array}\right.
\label{eq:coeffiecents}
\end{equation}
where $\omega_{\pm}=1\pm\Gamma_{+}\Gamma_{-}e^{i\phi}$, $\phi=\phi_{-}-\phi_{+}$ ($e^{i\phi}=\pm 1$ depending on $\theta$ and $\alpha$),
\begin{eqnarray*}
    && \Gamma_{\pm}=\frac{v_{\pm}}{u_{\pm}}=\frac{|\Delta_{\pm}|}{E+\Omega_{\pm}}=\frac{E-\Omega_{\pm}}{|\Delta_{\pm}|},\\
    && Z_{\sigma}=\frac{Z_{0}-\sigma Z_{1}\sin\theta}{\cos\theta},\quad
    Z_0=\frac{2mU_0}{k_F},\quad Z_1=2mU_1.\quad
\end{eqnarray*}
The dimensionless parameters $Z_0$ and $Z_1$ characterize the strengths of the purely potential and SO scattering, respectively.

Using the BTK formalism,\cite{BTK82} the normalized differential tunneling conductance is given by $G(E)=G_S(E)/G_N$, where
\begin{equation}
\label{GS-integrated}
    G_S(E)=\sum\limits_{\sigma}\int_{-\pi/2}^{\pi/2} d\theta\cos\theta\,G_{\sigma}(E,\theta),
\end{equation}
with the angle and spin resolved conductance given by
\begin{eqnarray}
\label{G-sigma}
    G_{\sigma}(E,\theta)&=&1+|a_{\sigma}(E,\theta)|^{2}-|b_{\sigma}(E,\theta)|^{2} \nonumber\\
    &=&\frac{1+|\Gamma_{+}|^{2}+Z_{\sigma}^{2}(1-|\Gamma_{+}\Gamma_{-}|^{2})/4}
    {|1+Z_{\sigma}^{2}(1-\Gamma_{+}\Gamma_{-}e^{i\phi})/4|^{2}},
\end{eqnarray}
and
\begin{equation}
\label{GN-integrated}
    G_{N}=\sum\limits_{\sigma}\int_{-\pi/2}^{\pi/2} d\theta\cos\theta\frac{1}{1+Z_{\sigma}^{2}/4}
\end{equation}
is the conductance for a normal metal/normal metal junction with the interface potential given by Eq. (\ref{interface-model}). We can see that the tunneling conductance depends on the incident spin orientation: $G_+(E,\theta)\neq G_-(E,\theta)$. In the absence of the interface SOC,
i.e. at $Z_{1}=0$, Eq. (\ref{G-sigma}) reduces to the known results [see, e.g., Eq. (23) of Ref. \onlinecite{KTKK96}].

The effect of the Andreev bound states is most pronounced in the low-transparency limit, i.e. when the interface barrier is so high that one can put $Z_\sigma\to\infty$. In this case, $G_N$ becomes small, but $G_{\sigma}(E,\theta)$ remains independent of the barrier height if $\Gamma_{+}\Gamma_{-}e^{i\phi}=1$. The last equation can be written in the form $\Delta_+/(E-\Omega_+)=\Delta_-/(E+\Omega_-)$, which has a zero-energy solution at $\alpha=\pi/4$ for all incident angles due to the fact that $\Delta_-=-\Delta_+$ (Ref. \onlinecite{Hu94}). Therefore, the normalized tunneling conductance at $E=0$ diverges, giving rise to a sharp ZBCP. In contrast, at $\alpha=0$ there are no zero-energy bound states near the interface and the tunneling probes the density of the quasiparticle states $N(E)$ in the bulk. In the $d$-wave case, at low energies the main contribution to the density of states comes from the vicinity of the gap nodes, yielding $N(E)\sim E$ (Ref. \onlinecite{Book}). This leads to a strong suppression of the tunneling conductance at low bias. We shall see below that the last conclusion surprizingly changes if the interface SOC is taken into account.

We use expressions (\ref{GS-integrated}) and (\ref{GN-integrated}) to calculate the normalized tunneling conductance for different values of the interface SOC at $\alpha=0$ and $\alpha=\pi/4$. For the potential barrier height we consider three cases: $Z_0=0$ (high transparency), $Z_0=1$ (medium transparency), and $Z_0=5$ (low transparency). The conductance is plotted as a function of the dimensionless excitation energy
$E/\Delta_{\textrm{0}}$.

Figures 2 and 3 show the tunneling conductance for $\alpha=0$ and $\alpha=\pi/4$, respectively, in the high and medium transparency cases.
We note that if $Z_{0}=Z_{1}=0$, then there is only Andreev reflection and $G(E)$ is nearly independent on $\alpha$,
monotonically decreasing from 2 as $E$ increases. This changes in the presence of the interface SOC, when a maximum appears in $G(E)$ in the subgap region at $\alpha=0$, as shown in the top panel of Fig.~2.

\begin{figure}
\includegraphics[width=7.5cm]{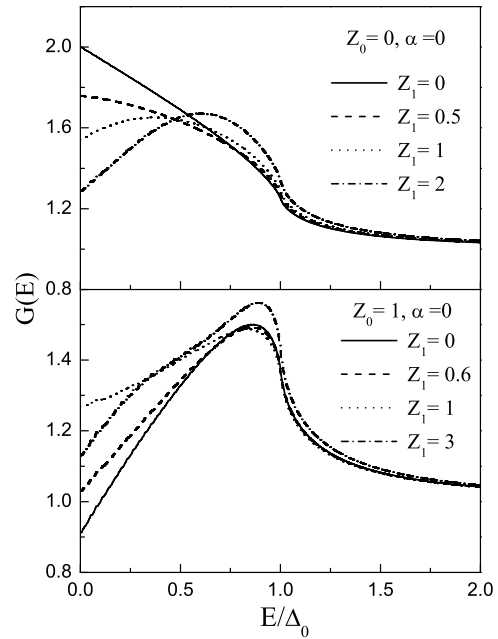}
\caption{The dimensionless tunneling conductance for $\alpha=0$ and different strengths of the
interface SOC, at $Z_{0}=0$ (top panel) and $Z_{0}=1$ (bottom panel).}
\end{figure}

\begin{figure}
\includegraphics[width=7.5cm]{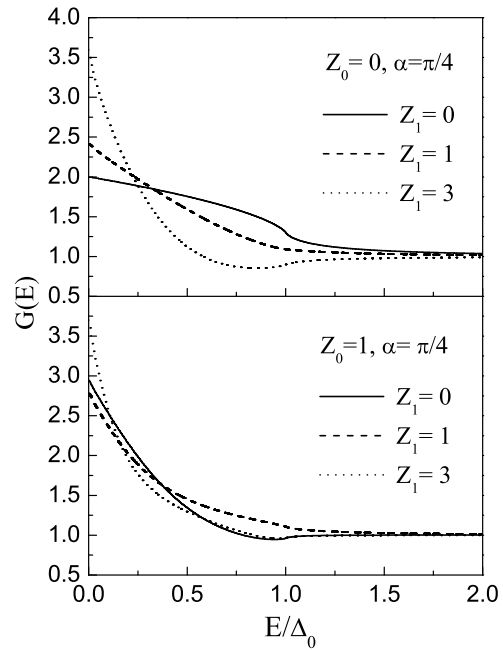}
\caption{The dimensionless tunneling conductance for $\alpha=\pi/4$ and different strengths of the
interface SOC, at $Z_{0}=0$ (top panel) and $Z_{0}=1$ (bottom panel).}
\end{figure}

Another significant feature of our results is that in the high-transparency case the effects of the SOC on the zero-bias conductance
are opposite for the two interface orientations: At $\alpha=0$ $G(0)$ is suppressed as $Z_{1}$ increases, while at $\alpha=\pi/4$ $G(0)$ is enhanced, as shown in the top panels of Figs.~2 and~3, respectively. To explain this we use the analytical expression
for the zero-bias conductance at $Z_0=0$:
$$
	G(0)=\frac{(x_1^2+1)(x_1^2-3)\arctan x_1+(x_1^2+3)x_1}{(x_2^2+1)\arctan x_2-x_2}\frac{x_2^3}{x_1^5},
$$
where $x_1=\sqrt{Z_1^2(1+e^{i\phi})/4-1}$ and $x_2=\sqrt{Z_1^2/4-1}$. At $Z_1=0$, we find $G(0)=2$.
At $\alpha=0$ ($\alpha=\pi/4$), we have $e^{i\phi}=1$ ($e^{i\phi}=-1$), and the above expression is a decreasing (increasing) function of $Z_1$.

In the medium-transparency case the interface SOC produces some enhancement of the tunneling conductivity, which is more pronounced in the $\alpha=0$ case.

Let us now discuss the case of a low transparency barrier, which is shown in Fig.~4. At $\alpha=\pi/4$, the interface SOC suppresses the height of the ZBCP.
At $\alpha=0$, there is no ZBCP, but the tunneling conductance at a small bias is enhanced by the SOC, reaching values of the order of 1 (in contrast, the peak at $E=\Delta_0$ is suppressed). To explain this enhancement we note that for $Z_1$ smaller than $Z_0$ we have $Z_\sigma\simeq Z_0\gg 1$. Then $G_S(0)\sim Z_0^{-4}$, while $G_N\sim Z_0^{-2}$, therefore, $G(0)\sim Z_0^{-2}\to 0$, similarly to the zero SOC case, see, e.g., Ref. \onlinecite{TanKash95prl}. However, if $Z_1\gtrsim Z_0$, then for some incident angles one has $|Z_\sigma|\ll 1$. It is the contribution from those angles that dominates the integrals in Eqs. (\ref{GS-integrated}) and (\ref{GN-integrated}), giving rise to $G_S(0),G_N\sim Z_1^{-1}$, therefore $G(0)\sim 1$.

\begin{figure}
\includegraphics[width=7.5cm]{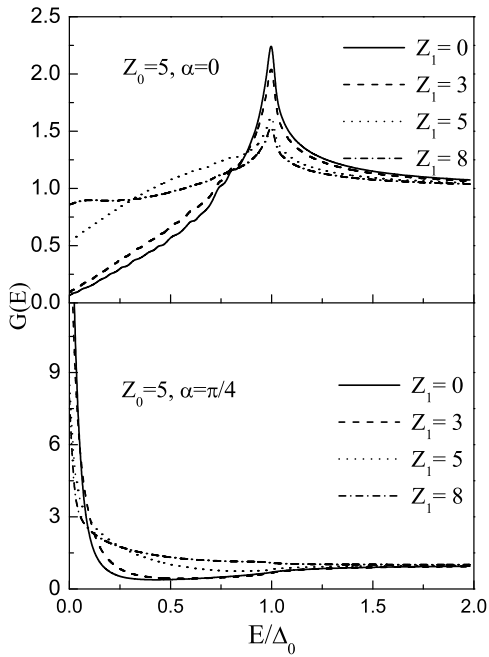}
\caption{The dimensionless tunneling conductance in the low transparency case, for different strengths of the
interface SOC, at $\alpha=0$ (top panel) and $\alpha=\pi/4$ (bottom panel).}
\end{figure}

To summarize, we have calculated the tunneling conductance of a junction between a normal
metal and a $d$-wave superconductor. Unlike the previous works, we take into account the SOC localized near the interface, which requires a modification
of the BTK formalism. We have shown that the interface SOC gives rise to several qualitative changes in the tunneling spectra.
The most prominent changes are as follows: In the case of a high-transparency junction, the normalized tunneling conductance at a small bias $E\ll\Delta_0$ is suppressed by the SOC for $\alpha=0$ and enhanced for $\alpha=\pi/4$. In the low-transparency junction, the trends are reversed, in particular, the zero-bias conductance at $\alpha=0$ is enhanced by the SOC.

This work was supported by a Discovery Grant from the Natural
Sciences and Engineering Research Council of Canada.

\end{document}